\documentclass{iau} 
\usepackage{graphicx}
\usepackage{natbib}
\usepackage{hyperref}
\usepackage{aas_macros}

\title[Close binaries and the abundance discrepancy] %% give here short title %%
{Close binary central stars and the abundance discrepancy - new extreme objects}

\author[Wesson et al.]{
R. Wesson,$^{1,2}$\thanks{E-mail: rw@nebulousresearch.org}
D. Jones,$^{3,4}$
J. Garc\'ia-Rojas$^{3,4}$,
R.L.M. Corradi,$^{3,4,5}$
and H.M.J. Boffin,$^{6}$}

\affiliation{
$^{1}$Department of Physics and Astronomy, University College London, Gower St, London WC1E 6BT, UK\\
$^{2}$European Southern Observatory, Alonso de C\'ordova 3107, Casilla 19001, Santiago, Chile\\
$^{3}$Instituto de Astrof\'isica de Canarias, E-38205 La Laguna, Tenerife, Spain\\
$^{4}$Departamento de Astrof\'isica, Universidad de La Laguna, E-38206 La Laguna, Tenerife, Spain\\
$^{5}$Gran Telescopio CANARIAS S.A., c/ Cuesta de San Jos\'e s/n, Bre\~na Baja, E-38712 Santa Cruz de Tenerife, Spain\\
$^{6}$European Southern Observatory, Karl-Schwarzschild-Str. 2, 85738 Garching bei M\"unchen, Germany
}

\pubyear{2017}
\volume{323}  %% insert here IAU Symposium No.
\jname{Planetary nebulae: multi-wavelength probes of stellar and galactic evolution}
\editors{X.-W. Liu, L. Stanghellino \& A. Karakas, eds.}
\begin{document}

\maketitle

\begin{abstract}
Recent work (\citealt{corradi2015}, \citealt{jones2016}) has shown that the phenomenon of extreme abundance discrepancies, where recombination line abundances exceed collisionally excited line abundances by factors of 10 or more, seem to be strongly associated with planetary nebulae with close binary central stars.  To further investigate, we have obtained spectra of a sample of nebulae with known close binary central stars, using FORS2 on the VLT, and we have discovered several new extreme abundance discrepancy objects.  We did not find any non-extreme discrepancies, suggesting that a very high fraction of nebulae with close binary central stars also have an extreme abundance discrepancy.

\keywords{planetary nebulae: general -- circumstellar matter -- stars: mass-loss -- stars: winds, outflows -- binaries: close -- ISM: abundances}
%% add here a maximum of 10 keywords, to be taken form the file <Keywords.txt>
\end{abstract}

\firstsection % if your document starts with a section,
              % remove some space above using this command.
\section{Introduction}

The abundances of heavy elements in photoionised nebulae may be determined from their strong, bright collisionally excited lines (CELs), whose emissivity depends exponentially on the temperature, or from the much weaker recombination lines (RLs), the emissivity of which has a weak power law dependence on the temperature.  Abundances derived from RLs and CELs do not agree, with the values from RLs exceeding those from CELs by a factor most commonly around 2--3, but reaching extreme values of 10 or more in about 10 per cent of planetary nebulae (PNe).  This abundance discrepancy problem has been known since the 1940s (\citealt{wyse1942}), but a full understanding of its causes remains elusive.  The abundance discrepancy factor ({\it adf}) for an ion is defined as the ratio of its RL abundance to its CEL abundance.  O$^{2+}$ is much the easiest ion to measure an {\it adf} for, being usually the most abundant heavy element, and having RLs and CELs in the optical.  In this article, unless otherwise specified, {\it adf} refers to {\it adf(O$^{2+}$)}.

A number of plausible mechanisms have been postulated which could account for the abundance discrepancy.  These include temperature fluctuations (\citealt{peimbert1967}), density fluctuations (\citealt{viegas1994}), hydrogen-deficient clumps (\citealt{liu2000}), X-ray illuminated quasi-neutral clumps (\citealt{ercolano2009}), and non-thermally distributed electrons (\citealt{nicholls2012}).  All except the hydrogen-deficient clump theory account for the discrepancy in the context of a chemically homogeneous nebula.  Since 2006 there have been suggestions that whatever the mechanism, it may be related to central star binarity.  Hf~2--2, one of the most extreme objects known with an abundance discrepancy of 80, has a binary central star with an orbital period so short that it must have undergone a common envelope (CE) phase (\citealt{liu2006}).  The difficulties in identifying such short period binaries meant that until recently, the numbers of planetary nebulae for which binarity was established and abundance discrepancy measured was small.  However, recently the picture has become clearer, with \citet{corradi2015} finding that three known post-CE binaries had nebulae with extreme abundance discrepancies, and then \citet{jones2016} strengthening the link with a study of the post-CE nebula NGC~6778, finding an abundance discrepancy of a factor of nearly 20 on average, and peaking at around 40 in the centre of the nebula.

We have obtained new spectra of a sample of PNe known to have close binary central stars.  The sample has revealed a number of objects with extreme abundance discrepancies, and no low-discrepancy objects.  There thus appears to be an almost perfect correlation between close binarity and extreme abundance discrepancies.

\section{Observations}

We obtained spectra of about 25 planetary nebulae using FORS2 mounted on UT1 (Antu) at the Very Large Telescope in Chile, in ESO programmes 093.D-0038(A) and 096.D-0080(A).  The spectra covered wavelengths from 3600-5000 and 5800-7200{\AA} at a resolution of 1.5{\AA}, sufficient to resolve recombination lines in the blended features at 4070{\AA} and 4650{\AA}.  The target nebulae were southern objects with known close binary central stars, for which there were no recombination line abundances in the literature.  The exception to that was Hf~2--2, known to have an extreme abundance discrepancy, which we included in our sample both as a control to verify our strategy and methodology, and with the hope that our spectra would be deep enough to permit a spatially resolved study.

The programme was designed as a filler to be carried out in almost any weather conditions, and so a number of objects in the sample were observed in less than ideal conditions, but several excellent spectra were obtained.   The data on NGC~6778, obtained early in the programme, revealed a new extreme object with an abundance discrepancy of $\sim$20, presented in \citet{jones2016}.  Upon the completion of the programme, we have detected recombination lines in seven further objects, including Hf~2--2.

\section{Analysis}

Spectra were reduced using standard \textsc{starlink} routines.  Cosmic rays were removed from the 2D frames using a combination of \textsc{starlink}'s figaro routines and a python implementation of the LAcosmic algorithm \citep{vandokkum2001}.  Emission line fluxes were measured using {\sc alfa} \citep{wesson2016a}, which we used in \citet{jones2016} to perform a similar analysis for the high {\it adf} PN NGC~6778. {\sc alfa} derives fluxes by optimising the parameters of Gaussian fits to line profiles using a genetic algorithm, after subtracting a globally-fitted continuum.

{\sc neat} \citep{wesson2012} was then used to derive final ionic and elemental abundances from these emission line fluxes.  The code corrects for interstellar extinction using the ratios of H$\gamma$ and H$\delta$ to H$\beta$ (the H$\alpha$/H$\beta$ ratio was not used to calculate the extinction but rather as a sanity check to ensure that line fluxes measured from the non-overlapping red and blue spectra were consistent) and the Galactic extinction law of \citet{howarth1983}; temperatures and densities are then derived from the standard diagnostics \citep[see][for full details]{wesson2012}.  Ionic abundances are then calculated from flux-weighted averages of the emission lines of each species using the previously derived temperatures and densities, and total abundances estimated using the ionisation correction scheme of \citet{delgado-inglada2014} The atomic data and ionisation correction functions used were as in \citet{jones2016}.

\section{Abundance discrepancies and evidence for H-deficient material}

We detected recombination lines in eight of our sample objects.  In all eight cases, the abundance discrepancy was extreme.  The values we obtained (including the Jones et al. 2016 value for NGC~6778) are listed in Table~\ref{adftable}.  Among the eight objects was Hf~2--2, for which we derive very similar results in the integrated spectrum to those of Liu et al. (2006).

\begin{table}
  \begin{tabular}{lcllll}
  {\bf Object} & {\bf O$^{2+}_{RLs}$ / O$^{2+}_{CELs}$} & T$_e$([O~{\sc iii}]) & T$_e$(BJ) & T$_e$(He~{\sc i}) & T$_e$(O~{\sc ii}) \\
 & \\
   Hf~2--2    & 80 & 8\,800 & 800 & 2\,000 & 2\,000 \\
   MPA 1759  & 80 & 11\,500 & -- & 4\,000 & -- \\
   Pe 1-9    & 75 & 10\,000 & -- & 4\,000 & -- \\
   NGC 6326  & 50 & 14\,500 & 8\,000 & 4\,000 & -- \\
   NGC 6337  & 30 & 12\,500 & -- & 3\,000 & $<$1\,000 \\
   NGC 6778 & 20 & 8\,800 & 4\,100 & 3\,000 & 1\,300 \\
   Fg 1       & 20 & 10\,450 & -- & 6\,000 & $<$1\,000 \\
   Hen 2--283 & 13 & 8\,800 & -- & -- & 3\,000 \\
  \end{tabular}
\caption{The abundance discrepancy for O$^{2+}$ and temperature diagnostics in the objects where recombination lines were detected}
\label{adftable}
\end{table}

We then estimated upper limits to the abundance discrepancy for the objects where recombination lines were not detected, using a spectral synthesis code to calculate the O$^{2+}$ recombination line spectrum, and varying the abundance until we found the value for which the RLs would have been detectable.  We found that the limits from our non-detections were not strong, and that an extreme abundance discrepancy could not be ruled out in any case.

Temperatures measured from various diagnostics support the picture of hydrogen-deficient material in these extreme objects.  The standard [O~{\sc iii}] line ratios give temperatures typical of a photoionised gas of `normal' composition - 8--12kK.  Three diagnostics give generally much lower temperatures: the Balmer jump (BJ), which lies close to the edge of our spectral coverage, but is detected with sufficient signal to noise in several nebulae; He~{\sc i} emission line ratios, and O~{\sc ii} recombination lines, which should probe the coldest and most metal-rich regions of the gas (\citealt{mcnabb2013}).  The temperatures implied by the various diagnostics are listed in Table~\ref{adftable}.

\section{Discussion}

The association between binarity and the abundance discrepancy suggests two hypotheses.  Firstly, that all PNe with a close binary central star have an extreme abundance discrepancy.  This work supports that hypothesis and almost doubles the number of objects known to have both a binary central star and an extreme abundance discrepancy.  However, there is currently thought to be at least one object which contradicts that.  NGC~5189 has a binary central star with a period of just over 4 days, but has an unusually low abundance discrepancy of 1.6 (\citealt{garcia-rojas2013}), albeit measured from a very small region of the nebula.  The Necklace is also a definite post-CE object, but no recombination lines were detected in deep spectra by \citet{corradi2011}.  We have recently obtained observations of NGC~5189 covering most of the nebula, from which we will be able to see if the low value found by \citet{garcia-rojas2013} is representative of the whole nebula or a chance observation of a low-{\it adf} region.  If these two objects really have a low {\it adf}, then in total we now have 14 of 16 extreme-{\it adf} objects in the sample of close binaries with known chemistry.

The complemenary hypothesis, that all nebulae with extreme abundance discrepancies have a close binary central star, is not yet well tested, but a number of extreme {\it adf} objects whose central star status is not yet confirmed would be fruitful grounds for investigation: Abell~30, Abell~58, NGC~1501, M~1-42 and M~3-32 are among the most extreme {\it adf} objects whose central star status is still unknown.  Spatially resolved studies of many high-{\it adf} objects have found that the RL abundances are strongly centrally peaked, further associating the phenomenon with the central star.  See e.g. \citet{liu2000}, \citet{jones2016} and \citet[ and this volume]{garcia-rojas2016}.

Whether the link between them is universal or not, it is nevertheless clear that central star binarity and nebular chemistry are strongly linked.  Any explanation for the most extreme abundance discrepancy must account for the fact that they are preferentially found in objects with close binary central stars.  The strong similarities long noted between high {\it adf} PNe and old nova shells such as CP Pup and DQ Her, which exhibit extremely low plasma temperatures and strong recombination lines, suggests that some kind of nova-like outburst from the close binary central star could be responsible for ejecting H-deficient material into the nebulae.  The $\sim$90\% hit rate for close binaries having an extreme abundance discrepancy would suggest that this eruption must happen soon after the formation of the main nebula.

\bibliographystyle{mnras}
\bibliography{references}

\begin{discussion}

%\discuss{Roberto M{\'e}ndez}{comment}
%\discuss{Dave Jones}{comment}

\end{discussion}

\end{document}